\documentstyle[preprint,aps]{revtex}

\begin{document}
\title{On the embedding of branes in five-dimensional spaces }
\author{F. Dahia$^{a}$ and C. Romero$^{b}$}
\address{$^{a}$Departamento de F\'{i}sica, Universidade Federal de Campina Grande,\\
58109-970, Campina Grande, Pb, Brazil\\
$^{b}$Departamento de F\'{i}sica, Universidade Federal da Para\'{i}ba, C.\\
Postal 5008, 58051-970 Jo\~{a}o Pessoa, Pb, Brazil\\
E-mail: fdahia@df.ufpb.br and cromero@fisica.ufpb.br}
\maketitle

\begin{abstract}
We investigate the embedding of \ $n$-dimensional branes in $(n+1)$%
-dimensional spaces. We firstly consider the case when the embedding space
is a vacuum bulk whose energy-momentum tensor consists of a Dirac delta
function with support in the brane. We then consider the embedding problem
in the context of Randall-Sundrum-type models, taking into account $Z_{2}$
symmetry and a cosmological constant. We employ the Campbell-Magaard theorem
to construct the embeddings and are led to the conclusion that the content
of energy-matter of the brane does not necessarily determine its curvature.
Finally, as an application to illustrate our results, we construct the
embedding of four-dimensional Minkowski spacetime filled with dust.
\end{abstract}

Higher-dimensional theories of gravity and, in particular, the recent
braneworld scenario\cite{add1,iadd,add2,rs1,rs2}, in which our visible
(under TeV scale of energy) spacetime appears as a hypersurface embedded in
a higher dimensional space, have greatly contributed to bring about a
renewed interest in both mathematical and physical aspects of embedding
theories of spacetime\cite
{ponce,wesson,coley,romero,overduim,book,sajko,maia,shiromizu,maarten,barcelo,ponce-equiv,mann,deruelle,ed,seahra}%
. As far as their mathematical structure is concerned embedding theories are
naturally subject to the theorems of differential geometry. These theorems,
in turn, may give us deeper insights on our understanding of the
relationship between the geometry of the lower-dimensional world and that of
the embedding space. In this respect, an important theorem due to Campbell
and Magaard, along with its extended versions, may shed some light on the
mathematical structure of five-dimensional non-compactified Kaluza-Klein
gravity theories \cite{book} as well as Randall-Sundrum (RS) models\cite
{rs1,rs2}. Local isometric embeddings of Riemannian manifolds have long been
studied in differential geometry. Of particular interest is a well known
theorem (Janet-Cartan)\cite{janet,cartan} which states that if the embedding
space is flat, then the minimum number of extra dimensions needed to
analytically embed a $n$-dimensional Riemannian manifold is $d$ , with $%
0\leq d\leq n(n-1)/2$. The novelty brought by Campbell-Magaard\cite
{campbell,magaard} theorem is that the number of extra dimensions falls
drastically to $d=1$ when the embedding manifold is allowed to be Ricci-flat
(instead of Riemann-flat). It is worth mentioning that the Campbell-Magaard
theorem can be extended to more general contexts, such as those in which the
higher-dimensional space is sourced by a cosmological constant or by a
scalar field and even by an arbitrary non-degenerate Ricci tensor\cite
{anderson,dahia1,dahia2,dahia3}. In this letter we are interested in the
case when the embedding higher-dimensional space has a source which is
singular and corresponds to a Dirac delta function whose support is the four
dimensional spacetime. Clearly, this question is motivated by the
Randall-Sundrum models, according to which the standard particle model is
confined to our four dimensional spacetime, and hence the matter-energy
content of the spacetime is represented by a delta term. Some aspects of
this problem have been studied recently \cite
{shiromizu,ponce-equiv,deruelle,seahra}, mainly in connection with the $Z_{2}
$-symmetry involved in the Randall-Sundrum models.

In this paper we are concerned with general aspects of the existence problem
of embeddings, in the same direction of the Campbell-Magaard theorem. More
specifically, we are interested in knowing what are the embedding conditions
when the energy-momentum tensor of the embedding space is singular. In other
words, we wonder if any $n$-dimensional spacetime with arbitrary metric and
energy-momentum tensor can be embedded in a $(n+1)$-dimensional vacuum
space. We also discuss the embedding problem in the context of
Randall-Sundrum type models.

Let us start our discussion with a brief review of the Campbell-Magaard
theorem, which states that any $n$\ dimensional space can be analytically
and isometrically embedded in a vacuum space of $n+1$ dimensions\cite
{campbell,magaard}. We recall that\ in the proof provided by Campbell and
Magaard the embedding problem is reduced to an initial value one. An outline
of their proof will be briefly given as follows. Consider the metric of the $%
n+1$ space written in a Gaussian coordinate system 
\begin{equation}
ds^{2}=\bar{g}_{ij}\left( x,y\right) dx^{i}dx^{j}+dy^{2},  \label{hds2}
\end{equation}
where $x=\left( x^{1},...,x^{n}\right) $, and Latin indices run from $1$ to $%
n$ while Greek ones go from $1$ to $n+1.$

It is easy to verify that the $(n+1)$-dimensional Einstein vacuum equations $%
R_{\mu \nu }=0$ expressed in the above coordinates are equivalent to the
following set of equations\footnote{%
In this letter we are adopting the following conventions for the Riemann and
Ricci tensors, respectively: $R_{\ \nu \alpha \beta \,}^{\mu }=\Gamma _{\nu
\alpha ,\beta }^{\mu }-\Gamma _{\nu \beta ,\alpha }^{\mu }+\Gamma _{\nu
\alpha }^{\tau }\Gamma _{\tau \beta }^{\mu }-\Gamma _{\nu \beta }^{\tau
}\Gamma _{\tau \alpha }^{\mu }$, $R_{\nu \beta }=R_{\ \nu \alpha \beta
}^{\alpha }$.}: 
\begin{eqnarray}
\bar{R}_{ij}+\left( \overline{K}_{ij}\overline{K}-2\overline{K}_{im}%
\overline{K}_{j}^{m}\right) -\frac{\partial \overline{K}_{ij}}{\partial y}
&=&0  \label{dyn} \\
\bar{\nabla}_{j}\left( \overline{K}^{ij}-\overline{g}^{ij}\overline{K}%
\right)  &=&0  \label{c1} \\
\bar{R}+\overline{K}^{2}-\overline{K}_{ij}\overline{K}^{ij} &=&0,  \label{c2}
\end{eqnarray}
where $\bar{\nabla}_{j}$ is the covariant derivative with respect to $\bar{g}%
_{ik}$; $\bar{R}_{ik}$, $\overline{R}$ , and $\overline{K}_{ik}$\ denote,
respectively, the Ricci tensor, the scalar curvature and the extrinsic
curvature of the hypersurface $y=y_{0}=const$, which has an induced metric
given by $\bar{g}_{ij}(x,y_{0})$, and $\overline{K}$ is the trace of $%
\overline{K}_{ik}.$ Recall that in the Gaussian coordinates adopted the
extrinsic curvature assumes the simple form: 
\begin{equation}
\overline{K}_{ij}=-\frac{1}{2}\frac{\partial \bar{g}_{ij}}{\partial y}.
\label{ext}
\end{equation}

Owing to the Bianchi identities not all of the equations (\ref{dyn}), (\ref
{c1}) and (\ref{c2}) are independent. In fact, we can show that the first
one propagates the others in the sense that, if Eqs. (\ref{c1}) and (\ref{c2}%
) are satisfied in a hypersurface $y=0,$ for example, and equation (\ref{dyn}%
) is valid in an open set of the $(n+1)$-dimensional space, then Eqs. (\ref
{c1}) and (\ref{c2}) will be satisfied in a certain open set of the $(n+1)$%
-dimensional space. Therefore, it is sufficient to demand that the
constraint equations (\ref{c1}) and (\ref{c2}) be satisfied in the
hypersurface $y=0$ to guarantee that they will hold in a family of
hypersurfaces $y=const$.

Now, by simple algebraic manipulation equation (\ref{dyn}) can be put in the
canonical form 
\begin{equation}
\frac{\partial ^{2}\overline{g}_{ij}}{\partial y^{2}}=F_{ij}\left( \overline{%
g}_{lm},\frac{\partial \bar{g}_{lm}}{\partial y}\right) ,  \label{CK}
\end{equation}
where $F_{ij}$ are analytical functions of their arguments. Therefore,
according to the Cauchy-Kowalewski theorem\cite{ck}, there exists a unique
analytical solution $\bar{g}_{ik}\left( x,y\right) $ satisfying the
analytical initial conditions : 
\begin{eqnarray}
\bar{g}_{ij}\left( x,0\right) &=&g_{ij}\left( x\right)  \label{cig} \\
\left. \frac{\partial \bar{g}_{ij}}{\partial y}\right| _{y=0}
&=&-2K_{ij}\left( x\right) .  \label{cih}
\end{eqnarray}

From the perspective of the embedding problem these initial conditions
represent the metric and the extrinsic curvature of the hypersurface $y=0$,
whereas the solution of equation (\ref{dyn}) gives the metric of the $\left(
n+1\right) -$dimensional space. Thus, if we can guarantee that the
constraint equations always have a solution, whatever the given metric $%
g_{ik}$ is, then the theorem is proved, since the solution found $\bar{g}%
_{ij}\left( x,y\right) $ will satisfy $R_{\mu \nu }=0$. Clearly, the
embedding map is given by equation the $y=0$.

It turns out, as Magaard has demonstrated\cite{magaard}, that the constraint
equations always have a solution. Indeed, by simple counting operation we
can see that there are $n(n+1)/2$ unknown functions (the algebraic
independent elements of extrinsic curvature) and $n+1$ constraint equations,
since the metric $g_{ij}\left( x\right) $\ must be considered as a given
datum. For $n>2$, there are more variables than equations. Thus using
equation (\ref{c2}) to express one element of $K_{ij}$ in terms of the
others, Magaard has shown that equation (\ref{c1}) can be put in a canonical
form with respect to $n$ components of $K_{ij}$ conveniently chosen. And
then, once more, the Cauchy-Kowalewski theorem ensures the existence of the
solution. It is important to note that, as we have said before, the number
of variables is greater than the number of \ equations; in this sense, we
can say that there are $\left( n+1\right) \left( n/2-1\right) $ degrees of
freedom left over.

Now it happens that when the $(n+1)-$dimensional embedding space is sourced
by a distribution of matter concentrated in\ \ a $n-$dimensional
hypersurface (thin shell), then the extrinsic curvature of the hypersurface
is not continuous.\ Following two different (although equivalent \cite
{mansouri}) approaches, namely, the Israel-Darmois formalism \cite
{darmois,israel} and the distributional method\cite{lich,bruhat,taub}, it
can be shown that the discontinuity of the extrinsic curvature is
proportional to the energy-momentum tensor of the hypersurface. In Gaussian
coordinates, we have 
\begin{equation}
\left[ K_{ij}\right] =\kappa \left( S_{ij}+\frac{g_{ij}}{1-n}S\right) ,
\end{equation}
where $\kappa $ is the gravitational constant in $(n+1)$-dimensions, $\left[
K_{ij}\right] =$ $%
\mathrel{\mathop{\lim }\limits_{y\rightarrow 0^{+}}}%
$ $K_{ij}-%
\mathrel{\mathop{\lim }\limits_{y\rightarrow 0^{-}}}%
$ $K_{ij}$, $S_{ij}$ is energy-momentum tensor of the hypersurface and $S$
is its trace. (The above equation is known as Lanczos equation\cite{mansouri}%
.)

It is clear that when there is a discontinuity of the extrinsic curvature
two different embeddings, one for each side of the hypersurface, are
required. As we have mentioned earlier, the extrinsic curvature considered
as part of the initial conditions in the scheme outlined above has some
degrees of freedom left over. If $n=4,$ there are five completely
independent components of the extrinsic curvature tensor. Each choice of the
these components possibly determines different embedding spaces. The
question now arises whether we can select two sets of these independent
components in such a way that Lanczos equation may be satisfied for any $%
S_{ij}$ .

Suppose we want to embed a $n$-dimensional spacetime with a given metric $%
g_{ij}$ and a specific energy-momentum tensor $S_{ij}$ in a $(n+1)$%
-dimensional vacuum space. Let $g_{ij}^{-}\left( x,y\right) $ be the metric
of the embedding space determined by solving the dynamical equation (\ref
{dyn}) and imposing the initial conditions data: the metric of spacetime $%
g_{ij}$ and the extrinsic curvature $K_{ij}^{-}.$ Of course $g_{ij}$ and $%
K_{ij}^{-}$ must satisfy the constraint equations (\ref{c1}) \ and (\ref{c2}%
).

A new metric $g_{ij}^{+}\left( x,y\right) $ could be obtained by taking
another extrinsic curvature $K_{ij}^{+}$ as initial data. Based on the
Lanczos equations, let us define $K_{ij}^{+}$ by the expression 
\begin{equation}
K_{ij}^{+}=K_{ij}^{-}+\kappa \left( S_{ij}+\frac{g_{ij}}{1-n}S\right) 
\label{h+}
\end{equation}
In order to be eligible for initial condition, $K_{ij}^{+}$ must satisfy the
constraint equations. As a consequence, the following conditions must be
imposed on $S_{ik}$ and $K_{ij}^{-}:$%
\begin{eqnarray}
\nabla _{i}S_{j}^{i} &=&0  \label{Sg} \\
K_{ij}^{-}S^{ij} &=&-\frac{\kappa }{2}\left( S_{ij}S^{ij}+\frac{1}{1-n}%
S^{2}\right) .  \label{Sh}
\end{eqnarray}
These equations imply that $S_{ij}$ and $g_{ij}$ are not completely
independent. The first equation, which does not involve $K_{ij}^{-}$,
requires that $S_{ij}$ be conserved with respect to the spacetime metric $%
g_{ij}$, which seems to be a reasonable condition. The second equation
establishes an additional constraint equation for the initial condition $%
K_{ij}^{-}$. As the extrinsic curvature has extra degrees of freedom ( five,
in the case of $n=4$ ),  this new constraint can, in principle, be solved
for any $g_{ij}$ and $S_{ij}$ chosen. We should emphasize here that $%
g_{ij}^{-}$ and $g_{ij}^{+}$ are analytical vacuum metrics defined in a
five-dimensional open set surrounding the spacetime. The metric $\overline{g}%
_{ij}$ of the embedding space will be given by a combination of the two
metrics $g_{ij}^{-}$ and $g_{ij}^{+}$ : 
\begin{equation}
\overline{g}_{ij}\left( x,y\right) =%
{g_{ij}^{+}\left( x,y\right) \qquad y\geq 0 \atopwithdelims\{. g_{ij}^{-}\left( x,y\right) \qquad y\leq 0.}%
\label{5g}
\end{equation}
The metric is continuous but its first derivative with respect to $y$ has a
discontinuity at $y=0.$ As the Cauchy-Kowalewski theorem ensures the
analyticity of the solutions $g_{ij}^{-}$ and $g_{ij}^{+}$, it follows that $%
\overline{g}_{ij}$ is piecewise analytical. Although $g_{ij}^{-}$ and $%
g_{ij}^{+}$ are vacuum metrics, $\overline{g}_{ij}$ is generated by a Dirac
delta source, i.e. 
\[
G_{\mu \nu }=-\kappa \delta \left( y\right) S_{\mu \nu }
\]
where the components of $S_{\mu \nu }$ involving the extra coordinate are
null.

It is remarkable that in this scheme the content of energy of the spacetime
represented by the energy-momentum tensor $S_{ij}$ does not necessarily
determine the curvature of spacetime. Indeed, as we have seen, the only
required relation between $S_{ij}$ and $g_{ij}$ is the conservation
equation. This fact allows for very unconventional situations to take place.
As we will see next, it is possible to have Minkowski spacetime filled with
dust.

Consider the five-dimensional space metric given by 
\[
^{\left( 5\right) }ds^{2}=%
{-\left( a^{+}y+1\right) ^{2}dt^{2}+dl^{2}+dy^{2}\qquad y\geq 0 \atopwithdelims\{. -\left( 2a^{-}y+1\right) ^{-1}dt^{2}+\left( 2a^{-}y+1\right) dl^{2}+dy^{2}\qquad y\leq 0,}%
\]
where $dl^{2}$ denotes the Euclidean spatial interval and $a^{+}$, $a^{-}$
are constants. (Note that for $y\neq 0$, the above metric corresponds to
that of a vacuum space). Of course, the embedding map $y=0$ gives Minkowski
spacetime. But if we calculate the extrinsic curvature, it is easy to see
that $K_{ij}^{-}=-diag\left( a^{-},a^{-},a^{-},a^{-}\right) $ and that $%
K_{tt}^{+}=a^{+}$ is the only non-null term of $K_{ij}^{+}$. Now, if we take 
$a^{+}=a^{-}=\frac{\kappa }{3}\rho $, it follows, from Lanczos equation,
that $S_{tt}=\rho $ and the other components are null. This means that the
embedded Minkowski spacetime is full of dust with uniform density equal to $%
\rho $ .

It is well known that embeddings can be used as a mechanism of matter
generation. This constitutes perhaps a key point of the so-called
space-time-matter theory (STM), also referred to as induced-matter-theory 
\cite{book,wesson1,ponce1,chodos,lidsey}. Here, in contrast, we are in the
presence of an utterly new and perhaps unexpected situation, where the
matter confined to the brane does not necessarily curve the spacetime if the
latter is conceived as a four-dimensional hypersurface embedded in a
five-dimensional space endowed with a metrical \ structure which is not of
differential class $C^{1}$.

Let us now employ the scheme described above to discuss the embedding of the
spacetime in a Randall-Sundrum type model. In this case the $(n+1)$%
-dimensional space is characterized by two essential properties: it
possesses $Z_{2}$ symmetry and is sourced by a cosmological constant $%
\Lambda $ in the bulk. Thus, the Einstein tensor of the higher-dimensional
space should satisfy the equation 
\begin{equation}
G_{\mu \nu }=\Lambda g_{\mu \nu }-\kappa S_{\mu \nu }\delta \left( y\right) .
\label{eq-brane}
\end{equation}
The $Z_{2}$ symmetry implies that the metric must obey the condition 
\begin{equation}
g_{\mu \nu }\left( x,y\right) =g_{\mu \nu }\left( x,-y\right) ,  \label{z2}
\end{equation}
which imposes the following relation between the ``up'' and ``down''
extrinsic curvatures of the brane \qquad 
\begin{equation}
K_{ij}^{+}=-K_{ij}^{-}.  \label{k-k}
\end{equation}
As a consequence, the extrinsic curvature is now completely determined by
the Lanczos equations 
\begin{equation}
K_{ij}^{+}=-K_{ij}^{-}=\frac{\kappa }{2}\left( S_{ij}+\frac{g_{ij}}{1-n}%
S\right) .  \label{k-lanczos}
\end{equation}

The metrics $g_{\mu \nu }^{+}$and $g_{\mu \nu }^{-}$ (respectively
associated with $K_{ij}^{+}$ and $K_{ij}^{-}$ as far as the initial
conditions are concerned)\ should now satisfy the equation 
\begin{equation}
G_{\mu \nu }^{\pm }=\Lambda g_{\mu \nu }^{\pm }.  \label{einspace}
\end{equation}
Using the same procedure outlined above employed to analyze the vacuum
equations, we can show that the existence of $g_{\mu \nu }^{+}$ and $g_{\mu
\nu }^{-}$ is guaranteed by the Cauchy-Kowalewski theorem provided that the
extrinsic curvatures $K_{ij}^{+}$ and $K_{ij}^{-}$ and the spacetime metric $%
g_{ij}$ satisfy the new constraint equations\cite{dahia1}: 
\begin{eqnarray}
\nabla _{j}\left( K^{ij}-g^{ij}K\right)  &=&0 \\
R+K^{2}-K_{ij}K^{ij} &=&-2\Lambda .
\end{eqnarray}
Since the extrinsic curvature $K^{ij}$ is given by (\ref{k-lanczos}), these
constraint equations imply the following equations between $S_{ij}$ and $%
g_{ij}:$%
\begin{eqnarray}
\nabla _{i}S_{j}^{i} &=&0  \label{eq} \\
\frac{\kappa ^{2}}{4}\left( S_{ij}S^{ij}+\frac{S^{2}}{1-n}\right) 
&=&2\Lambda +R.  \label{eq2}
\end{eqnarray}
It is not difficult to realize that $g_{\mu \nu }^{+}$ and $g_{\mu \nu }^{-}$
satisfy the condition $g_{\mu \nu }^{+}\left( x,y\right) =g_{\mu \nu
}^{-}\left( x,-y\right) $. Indeed, assuming that $g_{\mu \nu }^{-}\left(
x,y\right) $ is a solution of equation (\ref{einspace}) associated with the
initial conditions (\ref{cig}),(\ref{cih}) for$\ g_{ij}$ and $K_{ij}^{-}$,
we can verify that $g_{\mu \nu }^{-}\left( x,-y\right) $ also satisfies (\ref
{einspace}). Of course, the initial conditions of the latter now refer to $%
g_{ij}$ and $-\left( K_{ij}^{-}\right) $. Since the Cauchy-Kowalewski
theorem states that the analytical solution satisfying the initial
conditions is unique, then we can conclude that $K_{ij}^{+}$, satisfying (%
\ref{k-k}), generates $g_{\mu \nu }^{+}\left( x,y\right) $ with the
mentioned property, i.e. $g_{\mu \nu }^{+}\left( x,y\right) =g_{\mu \nu
}^{-}\left( x,-y\right) $. Therefore, we conclude that the metric of the $%
(n+1)$-dimensional space defined by (\ref{5g}) automatically obeys the $Z_{2}
$ symmetry and that the existence of the embedding depends on the constraint
equation between $S_{ij}$ and $g_{ij}$. Each pair $\left(
g_{ij},S_{ij}\right) $ that satisfies the constraint equations (\ref{eq}), (%
\ref{eq2}) can be embedded in a five-dimensional space with the $Z_{2}$
symmetry generated by (\ref{eq-brane}). In this scenario, unless additional
restrictions are imposed on the bulk, the constraint equations (\ref{eq})
and (\ref{eq2})\ seems to represent the only connection between the geometry
and the matter configuration of the brane.

\section{Acknowledgement}

C. Romero thanks CNPq for financial support.


\begin{references}
\bibitem{add1}  Arkani-Hamed, N., Dimopoulos, S. and Dvali, G., Phys. Lett. 
{\bf B 429}, 263 (1998).

\bibitem{iadd}  Antoniadis, I., Arkani-Hamed, N., Dimopoulos, S. and Dvali,
G., Phys. Lett. {\bf B 436}, 257 (1998).

\bibitem{add2}  Arkani-Hamed, N., Dimopoulos, S. and Dvali, G., Phys. Rev. 
{\bf D59}, 086004 (1999).

\bibitem{rs1}  Randall., L. and Sundrum, R., Phys. Rev. Lett. {\bf 83}, 3370
(1999).

\bibitem{rs2}  Randall, L. and Sundrum, R., Phys. Rev. Lett. {\bf 83}, 4690
(1999).

\bibitem{ponce}  Wesson, P. and Ponce de Leon, J. Math. Phys.,{\bf \ 33},
3883 (1992).

\bibitem{wesson}  Wesson P et al, Int. J. Mod. Phys. A {\bf 11}, 3247 (1996).

\bibitem{coley}  Abolghasen, H., Coley, A. A. and MacMannus, D. J., J. Math.
Phys.,{\bf \ 37}, 361 (1996).

\bibitem{romero}  Romero, C., Tavakol, R. and Zalaletdinov, R., Gen. Rel.
Grav. {\bf 28}, 365 (1996).

\bibitem{overduim}  J. M. Overduin, P. S. Wesson, Phys. Rep. 283, 303 (1997).

\bibitem{book}  P. S. Wesson, Space-Time-Matter, World Scientific,
Singapore, (1999).

\bibitem{sajko}  Sajko, W. N., Int. J. Mod. Phys. D{\bf 9}, 445 (2000).

\bibitem{maia}  Maia, M. D., gr-qc/9512002.

\bibitem{shiromizu}  Shiromizu, T., Maeda K., Sasaki, M., Phys. Rev. D {\bf %
62} 024012 (2000).

\bibitem{maarten}  Maarten, R., Phys. Rev. D {\bf 62}, 084023 (2000).

\bibitem{barcelo}  Barcelo, C., Visser, M., JHEP 0010:019 (2000).

\bibitem{ponce-equiv}  Ponce de Leon, Mod. Phys. Lett. A {\bf 16}, 2291
(2001).

\bibitem{mann}  Mannheim, P. D., Phys. Rev. D {\bf 63}, 024018 (2001).

\bibitem{deruelle}  Deruelle, N., Katz, J., Phys. Rev. D {\bf 64}, 083515
(2001).

\bibitem{ed}  Maia, M. D., Monte, E., Phys. Lett. A {\bf 272}, 9 (2002).

\bibitem{seahra}  Seahra, S., Wesson, P. Class. Quantum Grav. {\bf 20}, 1321
(2003).

\bibitem{janet}  Janet, M., Ann. Soc. Polon. Math. {\bf 5} 38 (1926).

\bibitem{cartan}  Cartan, E., Ann. Soc. Polon. Math. {\bf 6} 1 (1927).

\bibitem{campbell}  Campbell, J. ``A Course of Differential Geometry''
(Oxford: Claredon, 1926).

\bibitem{magaard}  Magaard, L, ``Zur einbettung riemannscher Raume in
Einstein-Raume und konformeuclidische Raume'' (PhD Thesis, Kiel, 1963).

\bibitem{anderson}  Anderson, E. and Lidsey, J. E., Class. Quantum Grav. 
{\bf 18}, 4831 (2001).

\bibitem{dahia1}  Dahia, F. and Romero, C, J. Math. Phys., {\bf 43}, 11,
5804, (2002).{\bf \ }

\bibitem{dahia2}  Dahia, F and Romero, C., {\bf gr-qc/0111094. (}J. Math.
Phys., in press).

\bibitem{dahia3}  Dahia, F and Romero, C, J. Math. Phys., {\bf 43}, 6, 3097
(2002){\bf .}

\bibitem{ck}  Courant, R. and Hilbert, D., ``Methods of Mathematical
Physics'', {\bf II}, (John Wiley \& Sons, 1989).

\bibitem{mansouri}  Mansouri, R., Khorrami, M., J. Math. Phys. {\bf 37}, 11,
5672 (1996).

\bibitem{israel}  Israel, W. Il Nuovo Cimento {\bf XLIV B}, 1 (1966).

\bibitem{darmois}  Darmois, G., ``Memorial de Science Mathematiques'', {\bf %
XXV} (1927).

\bibitem{lich}  Lichnerowicz, A, C. R. Acad. Sc. Paris, {\bf 273}, 528
(1971).

\bibitem{bruhat}  Choquet-Bruhat, Y., Commun. Math. Phys., {\bf 12}, 16
(1969).

\bibitem{taub}  Taub, A. H., J. Math. Phys., {\bf 21}, 6, 1423 (1980).

\bibitem{wesson1}  Wesson, P.S., Gen. Rel. Grav, {\bf 16}, 193 (1984).

\bibitem{ponce1}  Ponce de Leon, Gen. Rel. Grav, {\bf 20}, 539 (1988).

\bibitem{chodos}  Chodos, A. and Detweiler, S., Phys. Rev., {\bf D21},
2167.(1980).

\bibitem{lidsey}  Lidsey, J., Romero, C., Tavakol, R., and Rippl, S., Class.
Quantum Grav. {\bf 14}, 865 (1997).
\end{references}
\end{document}